\documentclass[usenatbib]{mn2e}

\usepackage{graphicx}
\usepackage{amssymb}
\usepackage{amsmath}
\usepackage{color}
\usepackage{abbreviations}
\usepackage{placeins}

\newcommand{\be}{\begin{equation}}
\newcommand{\ee}{\end{equation}}

\newcommand{\fracp}[2]{\left(\frac{#1}{#2}\right)}

\newcommand{\beq}{\begin{equation}}
\newcommand{\eeq}{\end{equation}} 
\newcommand{\ind}[1]{{\rm #1}}

\graphicspath{ {./}{figures/} }

\begin{document}

\title[Rayleigh-Taylor filaments in the Crab nebula]
{Rayleigh-Taylor instability in Magnetohydrodynamic Simulations of the Crab Nebula}

\author[Porth et al.]{Oliver Porth$^{1,2}$\thanks{E-mail: o.porth@leeds.ac.uk (OP)}, 
Serguei S. Komissarov$^{1,3}$\thanks{E-mail: serguei@maths.leeds.ac.uk (SSK)}, 
Rony Keppens$^{2}$\\
$^{1}$Department of Applied Mathematics, The University of Leeds, Leeds, LS2 9JT \\
$^{2}$Centre for mathematical Plasma Astrophysics, Department of Mathematics, KU Leuven,
 Celestijnenlaan 200B, 3001 Leuven, Belgium\\
$^3$Department of Physics, Purdue University, 525 Northwestern Avenue, West Lafayette, IN 47907-2036 
 }

\date{Received/Accepted}
\maketitle
\begin{abstract} 
    In this paper we discuss the development of Rayleigh-Taylor
    filaments in axisymmetric simulations of Pulsar wind nebulae
    (PWN).  High-resolution adaptive mesh refinement
    magnetohydrodynamic (MHD) simulations are used to resolve the
    non-linear evolution of the instability.  The typical separation
    of filaments is mediated by the turbulent flow in the nebula and
    hierarchical growth of the filaments.  The strong magnetic
    dissipation and field-randomization found in recent global
    three-dimensional simulations of PWN suggests that magnetic
    tension is not strong enough to suppress the growth of RT
    filaments, in agreement with the observations of prominent
    filaments in the Crab nebula.  The long-term axisymmetric results
    presented here confirm this finding.
\end{abstract}


\section{Introduction}

The Crab Nebula is a product of the supernova explosion observed by Chinese and Arab 
astronomers in 1054. It belongs to the class 
of Pulsar Wind Nebulae (PWN), bubbles of non-thermal plasma inflated inside of  
expanding supernova ejecta by magnetized neutron stars produced during core-collapse 
stellar explosions.   
The spectacular network of line-emitting filaments of the Crab Nebula is one of its 
most remarkable features. In projection, these filaments appear to occupy more or less 
the same space as the more amorphous non-thermal optical and radio emission. 
Most individual filaments are small scale structures but some are much longer 
and appear to cross almost the entire nebula. Images obtained at different epochs 
reveal radial expansion of the filamentary network \citep{trimble-68}.   
The speed of the proper motion of the filaments increases towards the outer edge. 
The line-of sight speeds, obtained via spectroscopic observations, show the 
opposite trend \citep{lawr-95}, thus confirming the overall radial expansion of the 
network. Moreover, the narrow band images of the Crab Nebula show that the 
filaments with low line-of-sight speed avoid the central part of the nebula 
image \citep{lawr-95}.  This indicates that the Crab filaments do not penetrate 
the whole volume of the nebula, as otherwise low line-of-sight-speed emission would 
be seen there.  Instead, the filaments reside near its outer edge, where they occupy 
a thick shell of thickness  0.3-0.7~pc, which is about one third of the nebula 
radius \citep{clark-83,lawr-95}. 

Initially, it was thought that the filaments could be the debris of the 
stellar envelope produced during the supernova explosion. However, this 
interpretation is in conflict with the low total mass of the filaments,  
$2-5 M_{\sun}$ and their low speed, only $\le 1500$~km/s, resulting in total 
energy of the explosion which is well below the typical one for core-collapse 
supernovae \citep{hester-08}. In order to bring these values up towards the 
expectations, one has to assume that most of the supernova ejecta is not 
visible yet. Most probably, the ejecta size significantly exceeds that of the 
Crab nebula and remains invisible due to low density ISM and hence a weak forward 
shock, whereas the observed thermal emission of the nebula comes as a result of 
the interaction between the inner part of the ejecta and the relativistic wind of 
the Crab pulsar \citep{rees-gunn-74,kc84a}.  

The high pressure of the hot bubble inflated by the pulsar wind inside of the 
ejecta drives a shock wave into the cold ejecta, heating its plasma and making it 
visible. Indeed, deep images of the Crab Nebula in high-ionization
lines reveal a sharp outer edge, which can be identified with this shock 
\citep{GF-82,SH-97}. The non-thermal emission is generally confined within this edge, 
in agreement with this interpretation, although the edge is not seen in  
it in the north-west part of the nebula, there the radio emission seems to extend 
beyond the thermal one \citep[e.g.][]{V-84}. However, the cooling time of the 
post-shock gas and brightness of its emission is a strong function of the ejecta 
density and the observations may simply indicate lower ejecta density in the 
NW-direction \citep{SH-97}.    
           
The ejecta is much denser than the PWN bubble and provided the shock, and hence 
the contact discontinuity separating the shocked ejecta from the bubble, expands 
with increasing speed, this configuration is similar to the one where a heavy fluid 
is placed on top of a light one in gravitational field. The latter is known to be     
Rayleigh-Taylor (RT) unstable. During the non-linear phase of this instability, 
the heavy fluid forms fingers which stream downwards and the light fluid 
forms bubbles rising between the fingers. 
\citet{CG-75} proposed that this is the origin for the thermal filaments of 
the Crab Nebula. The possibility of acceleration is strongly supported by both the 
observations and the theoretical models of PWN.  Indeed, the estimates of the 
nebula age based on its observed size and expansion speed are significantly shorter 
compared to that based on the time of the supernova explosion, implying accelerated 
expansion \citep[e.g.][]{trimble-68,bietenh-91a}. Strengthening this conclusion, 
the self-similar model of PWN inflated inside the ejecta with density 
$\rho\propto r^{-\alpha}$ by a pulsar wind of constant power 
yields the shock speed 
\begin{align}
v_{sh}\propto t^{1/(5-\alpha)}
\label{eq:v_sh}
\end{align}
\citep{CF-92}.  

The RT instability has been a subject of many theoretical studies. 
The original problem, involving ideal incompressible semi-infinite fluids 
in slab geometry, has been expanded to study the role of other factors, such as 
viscosity, surface tension, different geometry, magnetic field etc.      
For the original problem, the linear theory of the RT instability gives 
the growth rate 

\begin{align}
\omega^2 =  A g k \,,
\label{eq:omega}
\end{align}
where $g$ is the gravitational acceleration 
and $k$ is the wavenumber of the perturbation and where we introduce the Atwood number
$A=(\rho_2-\rho_1)/(\rho_2+\rho_1)$, where $\rho_1$ and $\rho_2$ are the mass 
densities of light and heavy fluids respectively. Thus, smaller scale structures 
grow faster. The transition to the non-linear regime occurs when the amplitude of 
the interface distortion becomes comparable to the wavelength. At the onset of 
this phase, the light fluid forms bubbles/columns of diameter $\sim\lambda$ 
which steadily rise with the speed 

\begin{align}
v_b\simeq 0.5\sqrt{g\lambda} \,,
\label{eq:v_b}
\end{align}
whereas the heavy fluid forms thin fingers approaching the state of free fall 
\citep[e.g.][]{DS-50,F-54,youngs-84,K-91,rama-12}. Thus at this stage, bubbles of larger 
scales grow faster and eventually dominate the smaller ones. This has been  
observed both in laboratory and in simulations \citep[e.g.][]{youngs-84,jun-95,SG-07}.
Interestingly, the initially dominating small scale perturbations appear 
to be washed out completely when much larger scales begin to dominate. 
Even if the initial spectrum of linear perturbations has a high wavelength cutoff, 
structures on the length scale exceeding it may appear via a kind of inverse 
cascade process, where smaller bubbles merge and create larger ones 
\citep{sharp-84,youngs-84}.        
The dynamics of bubbles and fingers is influenced by secondary Kelvin-Helmholtz 
instability, which facilitates transition to turbulence and mixing between the fluids.
This could be the reason for the observed disappearance of smaller scales.

The geometry of PWN is very different from that of the original RT problem. 
For example, the finite extension of PWN puts a natural upper limit on the 
length scale of RT-perturbations which may develop and the shell of heavy fluid 
is not thick compared to the observed size of RT fingers and bubbles.  
Moreover, the shell is bounded by a shock wave and the whole configuration is 
expanding, including the perturbations. 
\citet{vishn-83} studied linear stability of thin shells formed 
behind spherical shocks in interstellar medium. He found that for the ratio of specific heats
$\gamma<1.3$, the shell (and the shock) may experience unstable oscillations (overstability) 
for wavelengths below the shock radius. In geometric terms, the shell becomes rippled, extending 
further out in some places and lagging behind in others. For accelerated expansion, 
the RT instability is apparently recovered in the limit of planar shock. 
\citet{CF-92} applied the thin shell approach of \citet{vishn-83} to PWN. 
In particular, they found that, in agreement with the earlier finding 
by \citet{vishn-83},  in the spherical geometry  
the law of perturbation growth changes from exponential to power one and that only 
spherical harmonics of the degree $l\ge 5$ actually grow in amplitude.

\citet{jun-98} investigated the role of these factors in the non-linear regime via 
axisymmetric numerical non-relativistic hydrodynamic (HD) simulations. 
He considered the case of uniform supernova ejecta and isotropic pulsar wind of 
constant luminosity. In accordance with the expectations based on the linear thin-shell 
theory,  the results show rippling of the forward shock as well as developing of the 
RT fingers.   
They also show the gradual replacement of small-scale structures with larger ones, both in 
terms of linear and angular scales, similar to that seen in the earlier numerical 
studies of RT instability (see their Figure 6)\footnote{ \citet{jun-98} gives no 
information on the type and spectrum of initial perturbations.}. 
By the time of 4000~yr,  the dominant angular scale of the RT bubbles is 
about $\theta\sim \pi/20$ and the RT fingers have approximately the same linear size as 
the ripples.  The RT fingers are remarkably thin and coherent and reminiscent at least of 
some of the Crab filaments. However, at the current age of the Crab Nebula, $\sim 1000\,$yr,
the thickness of the mixing layer occupied by the fingers is much smaller, only 
approximately 1/15 of the PWN radius. This is about five times below the observed 
thickness of the Crab's filamentary shell. A similar discrepancy is found with respect 
to the scale of the shock ripples\footnote{Although \citet{jun-98} does not 
attempt to compare results with the Crab Nebula because of the idealized nature 
of the simulations, the parameters of the setup are actually based on Crab data.}.

The PWN plasma is magnetized and this motivates to investigate the role of magnetic field 
in the RT-instability. Since the magnetic field of the supernova ejecta is expected to be 
much weaker, the only relevant case is where the magnetic field is present only in the light
fluid and hence runs parallel to the interface.
Introduction of such field to the original RT problem leads to the growth rate 

\begin{align}
\omega^2 = A g k -\frac{(\mathbf{B\cdot
    k})^2/2\pi}{\rho_2+\rho_1} \, ,
\label{eq:omega_m}
\end{align}
where $\rho_1$ and $\rho_2$ are the mass densities of light and heavy fluids 
respectively and $\mathbf{k}$ is the wave-vector of the perturbation 
\citep{chandra-stability}. For the modes normal to the magnetic field, the 
growth rate of the non-magnetic case is recovered. For modes parallel to the 
field the magnetic tension suppresses the perturbations with wavelengths below the 
critical one, 

\begin{align}
\lambda_c = \frac{B^2}{g(\rho_2-\rho_1)} \, 
\label{eq:lambda_c}
\end{align}   
and the wavelength of fastest growing modes exceeds $\lambda_c$ by a factor of two.
2D and 3D computer simulations confirm these conclusions of 
the linear theory \citep{jun-95,SG-07}.  They also demonstrated that even 
magnetic field which is 
relatively weak compared to the critical one  may have significant 
effect on the non-linear evolution of RT fingers via inhibiting the development 
of secondary KH instability, thus leading to longer fingers.     

\citet{hester-96} applied the theory of magnetic RT instability to the Crab Nebula. 
Their key assumption was that the smallest structures of the Crab's filamentary 
network reminiscent of the RT bubbles and fingers had the wavelength of $2\lambda_c$, 
in the limit $\rho_2\gg\rho_1$. Using the observational 
estimates of density they found that ``the ends meet'' when the magnetic field 
strength is near the equipartition value based on the non-thermal emission.   
Such strong magnetic field is indeed expected near the interface in the 1D model 
of PWN by \citet{kc84a}.  However, there are several reasons to doubt this 
analysis. First, the multi-dimensional relativistic MHD 
simulations of PWN of recent years have demonstrated that many results of the 1D 
model on the structure and dynamics are incorrect. Secondly, the magnetic field 
does not suppress  modes normal to the magnetic field.  Finally, the gradual progression 
to larger scales at the non-linear phase, as described above, seems to make the task of 
identifying structures corresponding to the fastest growing linear modes virtually 
impossible.

In the context of PWN the interface acceleration is not an arbitrary parameter, 
but relates dynamically to the PWN pressure and the ejecta density.   
\cite{bucc-04} utilized the self-similar model of PWN evolution by 
\citet{CF-92} to derive the critical angular scale of magnetic RT instability. 
In the case of constant wind power, they obtained

\begin{align}
\theta_c/\pi = 8 \frac{P_m}{P_{tot}} \Delta f(\alpha)\, ,
\label{eq:theta_c}
\end{align}
where $P_m$ and $P_{tot}$ are the magnetic and total pressure of the PWN 
near the interface, $\Delta$ is the thickness of the shocked ejecta, 
$\alpha$ is the index of the ejecta density distribution $\rho\propto r^{-\alpha}$, 
and $ f(\alpha) = 1+ (3-\alpha)/(6-\alpha)$. For uniform ejecta ($\alpha=0$) in the
adiabatic case, one finds $\Delta \simeq 0.02$ \citep{jun-98}, and hence 
$ \theta_c/\pi \simeq 0.25 ({P_m}/P_{tot})\, . $
One can see that for magnetic field of equipartition strength, the critical 
scale is getting close to $\pi$, implying full suppression of the RT 
instability along the magnetic field. 
To test this result, \cite{bucc-04} carried out 2D 
relativistic MHD simulations intended to study the dynamics in the 
equatorial plane of PWN\footnote{ Like in the model of \citet{kc84a}, the symmetry 
condition prohibits motion in the polar direction.}.  They considered equatorial 
sections of angular size up to $\pi/6$ and employed periodic boundary conditions 
in the azimuthal direction. The 1D model of \citet{kc84a}, with its purely 
azimuthal magnetic field was used  to setup the initial solution and 
the boundary conditions in the radial direction. The results of these simulations 
generally agreed with Eq.~(\ref{eq:theta_c}), demonstrating suppression of RT 
instability in models where the magnetic fields builds up to 
the equipartition value near the interface with the shocked ejecta.  
This conclusion is in conflict with the analysis of \citet{hester-96} 
who identify $\lambda_c$ with structures as small as $1\farcs5$ in the 
sky, which corresponds to $\theta_c\sim \pi/300$, and yet deduce a magnetic field 
of equipartition strength. 

The discovery of the highly non-spherical ``jet-torus'' feature in the inner 
part of the Crab Nebula \citep{weiss-00}, and subsequent theoretical and computational 
attempts to understand the origin of this feature have lead to a dramatic revision of 
the Kennel-Coroniti model \citep[e.g.][]{bogovalov-khan-02b,lyub-02,ssk-lyub-03,
ssk-lyub-04,delzanna-04,bogovalov-05,camus-09,porth-13,porth-14}. 
The KC-model describes the flow inside PWN as laminar radial expansion whose 
speed gradually decreases from its highest value just downstream  of the 
pulsar wind termination shock to its lowest value at the interface with the 
supernova ejecta. This deceleration is accompanied by a gradual amplification 
of the purely azimuthal magnetic field from its lowest value at the termination 
shock to its highest value at midpoint where the magnetic pressure is approximately 
equal to that of particles.  In reality, both the termination shock and the 
flow downstream of this shock are highly non-spherical with strong shears. 
The termination shock is highly unsteady and the motion inside the nebula 
is highly turbulent. The magnetic field of 
PWN is strongest not near its edge but at its center. While the KC-model requires 
the pulsar wind to be particle-dominated, which is in conflict with the theory of 
pulsar winds, our recent results show that the complex 3D dynamics of 
PWN allows the wind to be Poynting-dominated \citep{porth-13,porth-14}.     

These global simulations have reproduced many of the observed features of the inner 
Crab Nebula, which was their main objective. However, they failed to capture the  
development of thermal filaments.  
Only during our latest study \citep[see arXiv preprint of][]{porth-14}, 
we noticed what looked like ``embryos'' of RT fingers. In order to check this, we 
continued our reference 2D simulations all the way up to the 
current age of the Crab Nebula, by which time these embryos turned into fully 
developed structures. 
Unfortunately, we could not (yet) do the same in our 3D simulations due to their 
prohibitively high cost. 
In the present paper, we describe in details this part of our study together with 
the additional simulations carried out mainly to investigate the role of numerical 
resolution.

\section{Simulations overview} \label{sec:simulations}

The numerical method, the use of adaptive grid, as well as the initial and boundary conditions 
of the numerical models presented here are exactly the same as described in 
\citet{porth-14}. For this reason, we describe here only few key features and refer 
interested readers to that paper for details.  
The simulations have been carried out with the adaptive grid code 
MPI-AMRVAC \citep{amrvac}, using the module for integrating equations 
of ideal special relativistic MHD. The scheme is third order accurate on smooth solutions. 
Outside of the termination shock region,  we use an HLLC solver \citep{Honkkila:2007:HSI:1232960.1233238}, which significantly 
reduces numerical diffusion compared to the normal HLL solver. 
We employ cylindrical coordinates and used cells of equal sizes along the $z$ and $r$ 
coordinates. The base level of AMR includes $64\times32$ cells. 
From three to six more levels are used to resolve the PWN, depending on the model, 
and even more levels are introduced to fully resolve the pulsar wind and its termination 
shock. When we study the influence of 
numerical resolution on the development of the RT instability, we only change the number of allowed 
grid levels in the PWN zone, while keeping the same number of levels in the pulsar wind zone. 
Thus in the model with the lowest resolution (model A0, see Table~\ref{tab:simulations}), 
the relevant effective grid size is  $512\times256$ cells, whereas in the highest resolution 
model (A3) it is $4096\times2048$ cells. When the solution is scaled to the size of the Crab 
Nebula, the cell-size equals to $\Delta x=3.9\times 10^{16} \rm cm$ in the model A0
and $\Delta x=4.9\times 10^{15} \rm cm$ in the model A3.  

Initially, the computational domain is split in two zones separated by
a spherical boundary of radius $r_i=10^{18}$cm. The
outer zone describes a radially-expanding cold supernova ejecta. The 
ejecta is described as a radial flow with constant mass density $\rho=\rho_\ind{e}$ 
and the Hubble velocity profile $ v=v_\ind{i} (r/r_\ind{i})$. This is suitable 
for such young PWN like the Crab Nebula. The values of $\rho_\ind{e}$ and $v_\ind{i}$ 
are determined by the condition that the total mass and kinetic energy of the ejecta 
within $r_\ind{i}<r<5r_\ind{i}$ are $3M_{\sun}$ and $10^{51}$erg respectively.  
The inner zone is filled with the unshocked pulsar wind. 
To monitor the mass-fractions of PWN versus SNR material, we solve an additional conservation law
\begin{align}
  \frac{\partial}{\partial t} (\Gamma\rho\tau) + \nabla_i(\Gamma \rho\tau v^i) = 0
\end{align}
and inject $\tau=1$ with the PWN while $\tau=0$ elsewhere.  Hence we have $\rho_{\rm PW}=\tau \rho$ for the (leptonic) material injected with the PWN and $\rho_{\rm SNR}=\rho(1-\tau)$ for material originating in the SNR.  

\begin{table}
\begin{center}
\caption{Simulation parameters.  ID - the model name, $\sigma_0$ - the magnetization of the 
pulsar wind, effective Grid size - relevant for the nebula, $\Delta x$ - the cell size in the nebula in units of $10^{16}\rm cm$}
\begin{tabular}{@{}lllll}
ID &  $\sigma_{0}$  & Grid size & $\Delta x$ \\
\hline
\hline
A0   & 0.01 & $256\times512$  & 3.90 \\
A1   & 0.01 & $512\times1024$ & 1.95 \\
A2   & 0.01 & $1024\times2048$ & 0.98\\
A3   & 0.01 & $2048\times4096$ & 0.49\\
B1   & 1.00 & $512\times1024$ & 1.95 \\
\end{tabular}
\end{center}
\label{tab:simulations}
\end{table}

The angular distribution of the wind power is based on the monopole model 
of \citet{michel-73}, where it varies with the polar angle as $\propto\sin^2\theta$. 
Following \citet{bogovalov-99}, we use the split-monopole approximation to 
introduce the stripe-wind zone corresponding to the oblique dipole with 
the magnetic inclination angle $\alpha$. In all models we put $\alpha=45^{\circ}$, 
the value preferred in the model of the Crab pulsar gamma-ray emission by \citet{harding-08}.    
Most models in this study have the rather low wind magnetization parameter $\sigma_0=0.01$.
This is because 2D models with significantly higher magnetization develop artificially 
strong polar outflow. However, as we demonstrate here by including model B1 with $\sigma_0=1$, this does not make a noticeable 
impact on the development of the RT instability away from the poles.
The wind Lorentz factor is set to 
$\Gamma=10$ and the total wind power to the current spin-down power of the Crab pulsar, 
$L=5\times10^{38} \rm erg\,s^{-1}$ \cite[e.g.][and references therein]{hester-08}.     
The spin-down time of the Crab pulsar is about 700~yr \citep{LPS-93}, which is below 
the age of the Crab Nebula, $\tau_\ind{sp}\sim 960\,$yr, and future attempts of more accurate 
modelling of the nebula should take this into account.

\begin{figure*}
\begin{center}
\includegraphics[width=60mm]{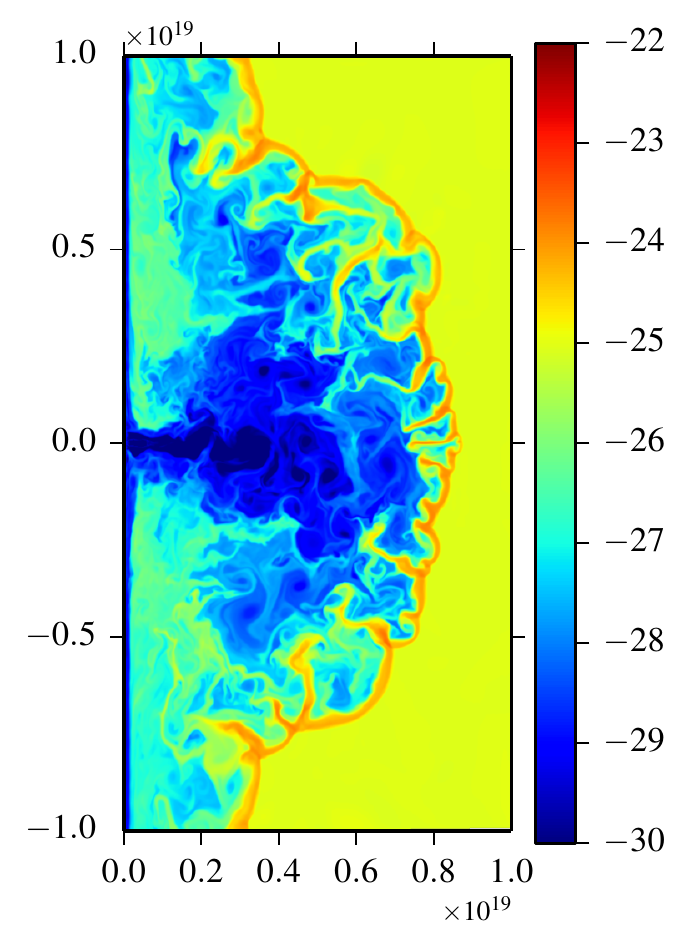}
\includegraphics[width=60mm]{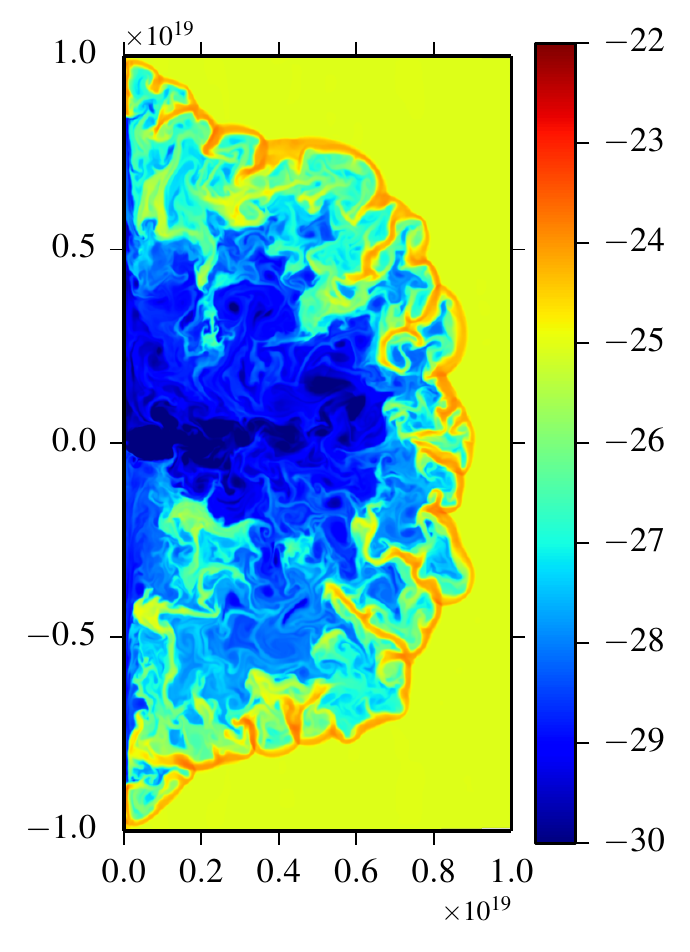}
\includegraphics[width=47mm]{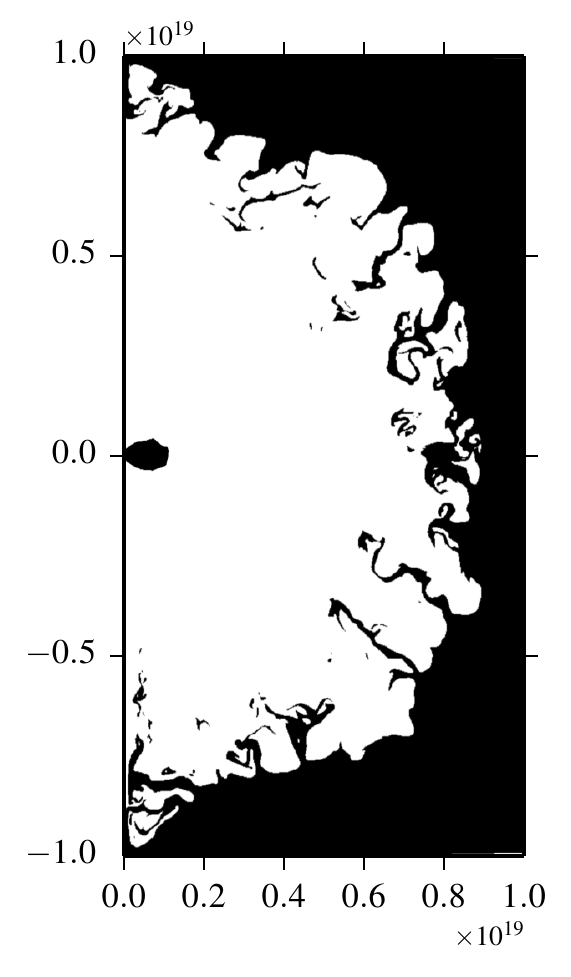}
\caption{Rayleigh-Taylor ``fingers'' in 2D simulations.  The left and
  middle panels show $\log_{10}\rho$ at $t\simeq 1060$ years in
  simulation runs B1 (left) and A1 (middle).  Note that in the low
  magnetization A1 run ($\sigma_0=0.01$) the jet is not able to penetrate
  far into the SNR.  The right-hand panel illustrates the way of
  determining the nebula radius $r_{\rm n}$.  It shows the mask used
  in order to separate PWN from the supernova shell.  The nebula
  radius is then obtained as the volume average $r_{\rm
    n}=\left(3V/4\pi\right)^{1/3}$.  The conspicuous ``fingers'' form
  via the Rayleigh-Taylor (RT) instability of the contact
  discontinuity between the PWN and the supernova shell.  The typical
  ``mushroom'' morphology of the RT-fingers, characteristic of the
  non-linear stages of this instability, is not seen here, most likely
  because of the interaction with the turbulent flow of PWN.  }
\label{fig:A1vsB1}
\end{center}
\end{figure*}

\section{Results}\label{sec:results}

One of the important results of \citet{porth-14} is that the global  
dynamics of axisymmetric 2D models and that of 3D models with strong 
magnetization of the pulsar wind differ dramatically. The 2D models develop
strong axial compression and produce powerful polar outflows. 
This results in a highly elongated shape of the nebula. This is in 
sharp contrast with the observations of the Crab Nebula, which is only 
moderately elongated. In contrast, the total pressure distribution of 
3D models is almost uniform and their shape remains approximately spherical.  
For 2D models to remain approximately spherical, the magnetization of the
pulsar wind should be low. From the perspective of studying the 
RT-instability in 2D, it looks like this makes us choose between two ``evils'' -- 
either to focus on the high-sigma models with their unrealistic overall 
geometry or on the low-sigma models with potentially weaker magnetic field 
in the nebula. Given the results of previous studies, the magnetic field 
strength can be important for development of the RT-instability 
\citep{jun-95,SG-07,bucc-04}. 
To clarify this issue we run two models, A1 and B1,  which differ only by 
the wind magnetization (both these models were studied in \citet{porth-14}). 
It turns out that the RT-instability yields very 
similar filamentary structure in these two cases everywhere apart from the 
polar zones, as one can see in figure~\ref{fig:A1vsB1}, which illustrates 
the solutions at the time $t\simeq1060\,$yr\footnote{The nebula age is given by 
the simulation time $t$ plus the initial time 
$t_0=r_{\rm i}/v_{\rm i}\simeq210~$years,
assuming initial expansion with constant $v_{\rm i}$.}. 

There are two main reasons behind this similarity of A1 and B1 models. 
First, in axial symmetry, the azimuthal magnetic field has no effect on the 
growth of RT perturbations as there is no mode-induced field line bending since $\mathbf{B}\cdot\mathbf{k}=0$. 
Second, the expansion rate of 
the nebula in the equatorial direction has not been altered dramatically 
in the high-sigma B1 model compared to the A1 one. Indeed, the equatorial 
radii are more or less the same in both models. Given this result, we 
decided to focus on the low sigma model in the rest of our study.

Figure \ref{fig:A1vsB1} shows a number of anticipated features. 
Similar to what was found in \cite{jun-98}, we see that 1) the initially
spherical shock front is now heavily perturbed and bulges out between
the RT fingers; 2) some of the filaments become detached from the shell; 
3) the filaments do not exhibit the ``mushroom caps'' characteristic 
of the single-mode simulations \citep{jun-95}. 
However, in contrast to \cite{jun-98}, we do not see significant density 
enhancements at the heads of the filaments. Moreover, the filaments 
extend much further into the nebula in our simulations, up to the distance 
of up to $1/4\,r_\ind{n}$, which is much closer to the value of $1/3\,r_\ind{n}$
deduced for the Crab Nebula. Visually, the scale of the shock ripples is also 
not that far away from the observed one.

Although these results looked very encouraging, it was not 
clear what exactly set the scale found in the simulations. In contrast 
to the previous studies we did not impose any perturbations of the shock 
front at the beginning of the runs. Instead, the RT mechanism amplified 
perturbations which had been imparted on the shock by the unsteady 
flow inside the PWN bubble. Visual inspection of Figure~\ref{fig:A1vsB1} 
hints that the scale of the dominant RT-modes could be related to the 
size of the termination shock, which sets the scale of large-scale eddies 
emitted by the shock into the PWN. On the other hand, numerical 
viscosity could also set the scale. 

As in any numerical study, it is imperative to check the resolution
dependence of our results.  Increased resolution leads to a reduction
of the numerical viscosity which in turn can influence the instability
growth.  The numerical viscosity of our third order reconstruction
scheme is expected to scale linearly with resolution, a behavior
established for example in high order WENO-type schemes
\citep{ZSSZ-03}.  The scaling of the viscous growth rate in the
Rayleigh-Taylor problem is well known \citep[e.g.][and references
therein]{K-91} and leads to

\begin{align}
k_{m}\propto \nu^{-2/3}
\end{align}
for the wave-number of the fastest growing mode. In terms of the
wavelength and cell size this reads as $\lambda_{\rm m}\propto \Delta x^{2/3}$. 
It is much more difficult to predict the outcome in the nonlinear regime because 
of the earlier saturation of small wavelength modes and the possible inverse 
cascade.   
  
\begin{figure*}
\begin{center}
\includegraphics[width=65mm]{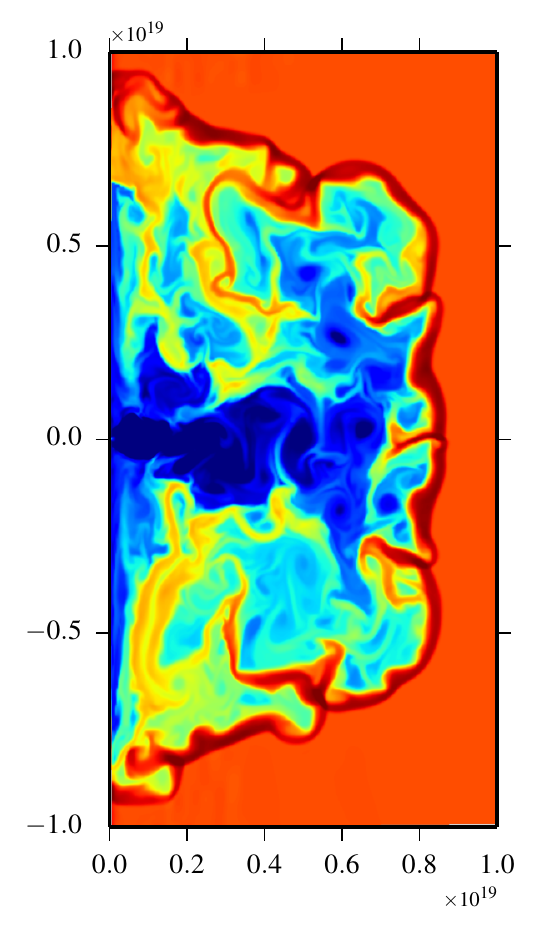}
\includegraphics[width=65mm]{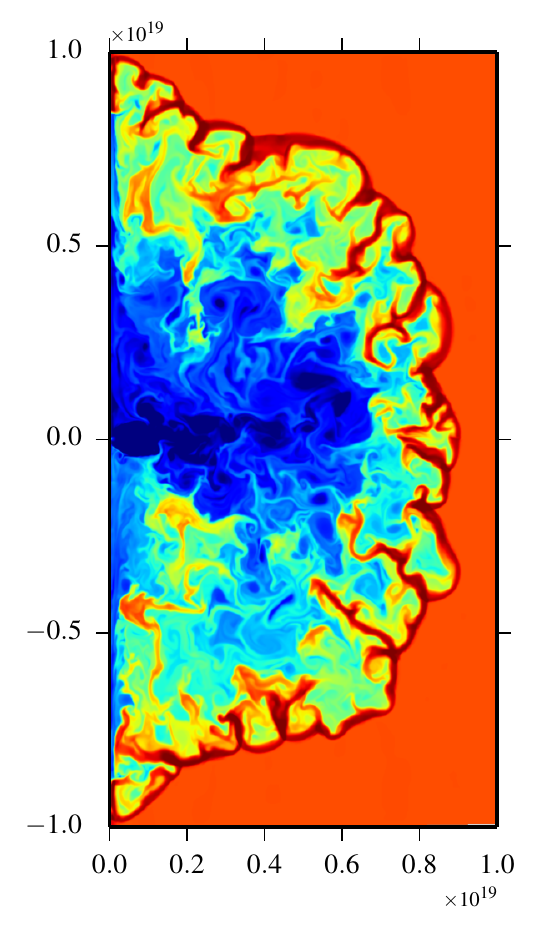}
\includegraphics[width=65mm]{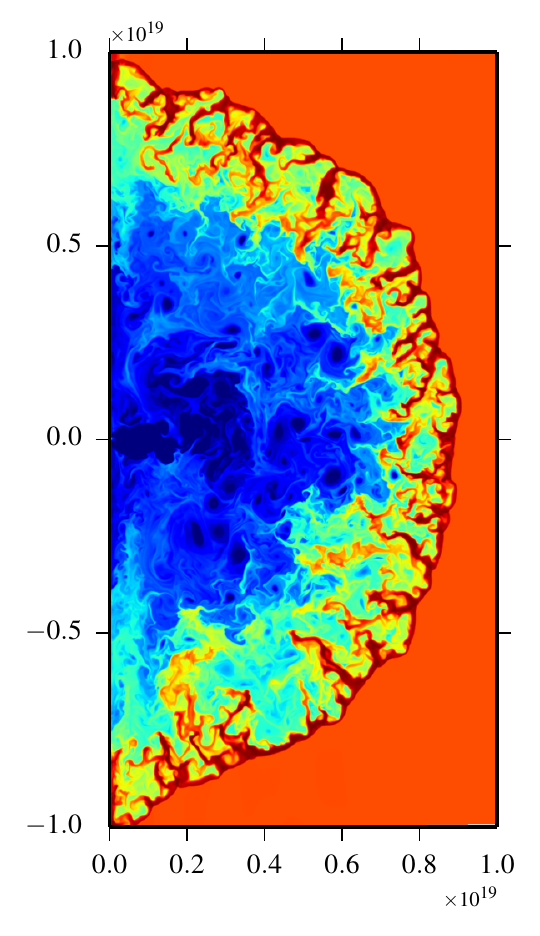}
\includegraphics[width=65mm]{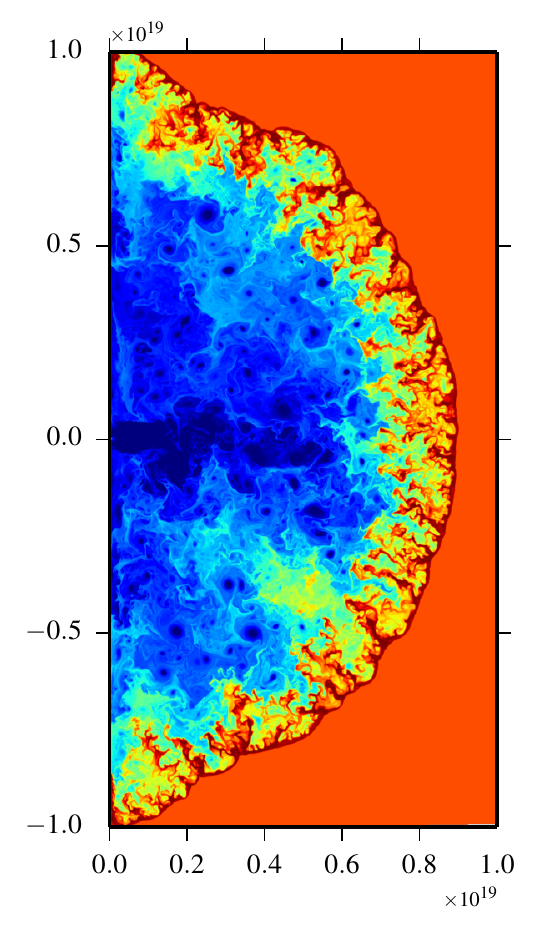}
\includegraphics[width=50mm]{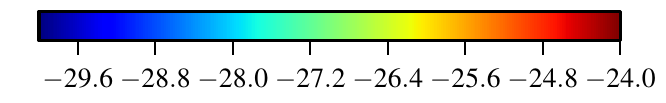}
\caption{Logarithmic densities showing the entire nebula and filaments at $t\simeq1060\,\rm years$
  with increasing resolution.  The models shown here are A0 (top left), 
A1 (top right), A2 (bottom left) and A3 (bottom right). 
Each next model has twice the resolution of the previous one.  }
\label{fig:resolution}
\end{center}
\end{figure*}

\begin{figure}
\begin{center}
\includegraphics[width=65mm]{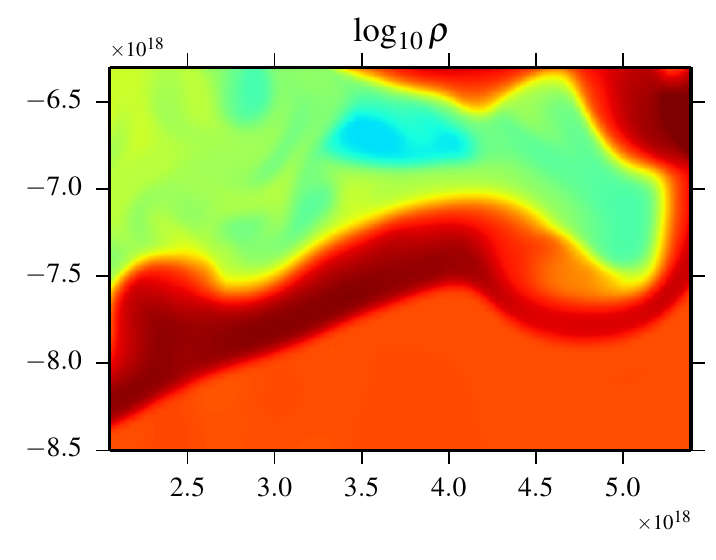}
\includegraphics[width=65mm]{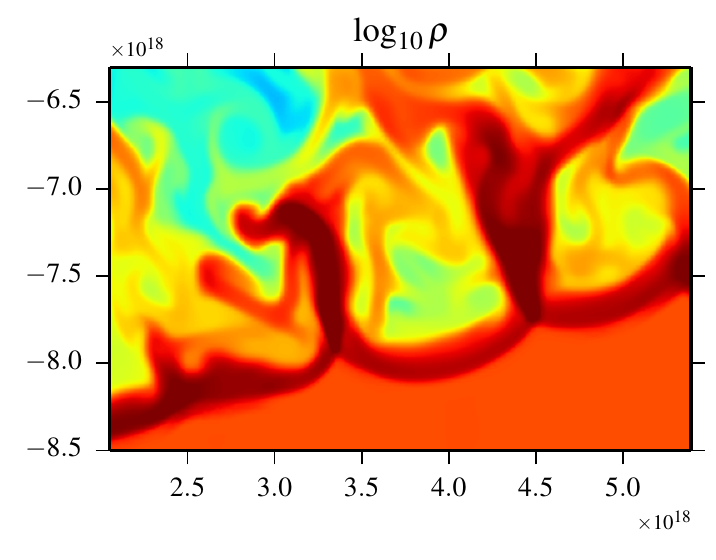}
\includegraphics[width=65mm]{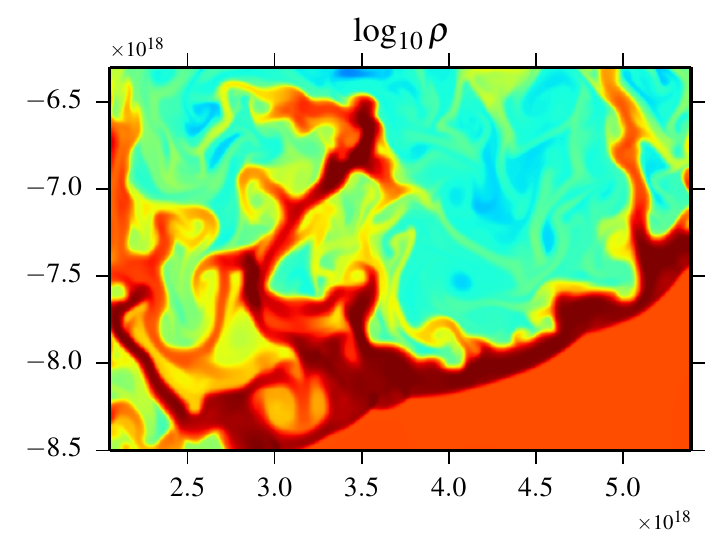}
\includegraphics[width=65mm]{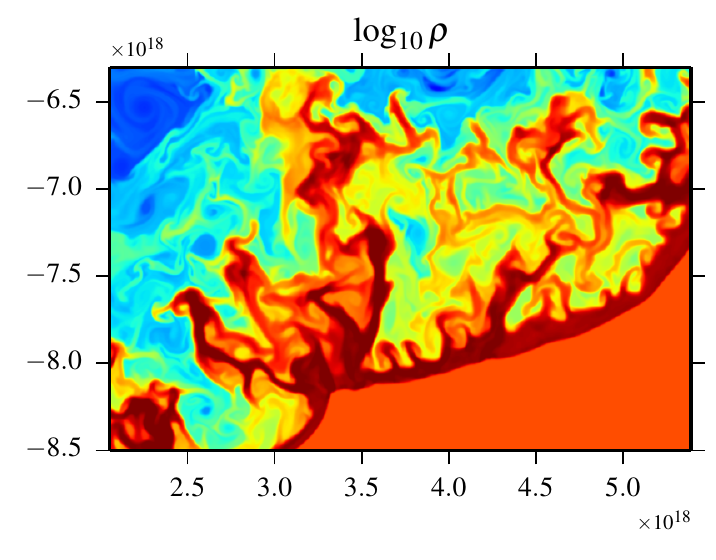}
\includegraphics[width=63mm]{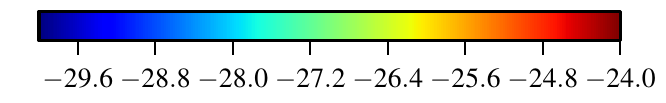}
\caption{Zoomed in views of logarithmic densities showing the southern
  shell and filaments with increasing resolution to illustrate the ``filament trees''.  Simulations A0,
  A1, A2 and A3 (top to bottom) are shown at
  $t\simeq 1060~\rm years$.}
\label{fig:zoom-rho}
\end{center}
\end{figure}

In order to study the role of resolution in the nonlinear regime, 
we run three more models A0, A2, and A3,  which differ 
from the A1 model only by the numerical resolution inside the PWN
bubble (see table~\ref{tab:simulations}). Figures \ref{fig:resolution} and 
\ref{fig:zoom-rho} show the density distribution found in these models.  One can see that 
while the size of the termination shock in all of them is more or less 
the same, with increasing resolution the power of RT features is progressively 
shifted towards smaller scales - the forward shock becomes more rounded and the 
RT-fingers become more numerous and small scale.       

In order to quantify the dominant scales, we analyze the 
surface mass density distribution defined via the integral 

\begin{align}
\Sigma(\theta) = \int_{0.6 r_{\rm n}}^{1.2 r_{\rm n}} \rho(r,\theta) r^2 dr \, .
\end{align}
Then we  
subtract the mean value,  $\Delta\Sigma(\theta)=\Sigma(\theta)-\bar{\Sigma}$, 
and use the Fourier decomposition to obtain the power spectrum $P(\Delta\Sigma)(m)$  
of the residual fluctuations.\footnote{Note that the integrand is also shown in figure \ref{fig:time_evol}.}  

\begin{figure*}
\begin{center}
\includegraphics[width=70mm]{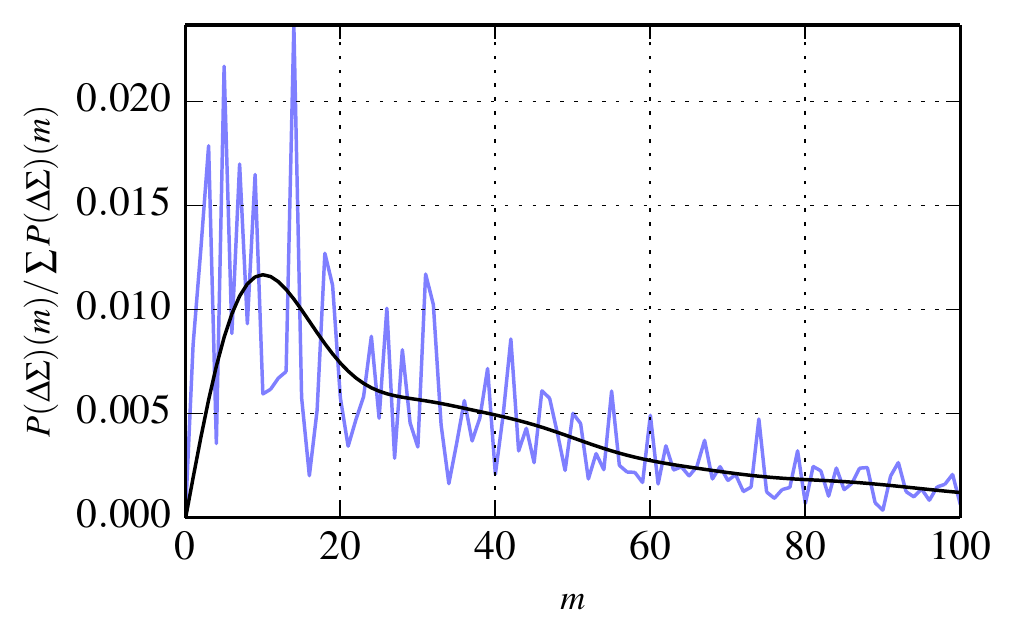}
\includegraphics[width=70mm]{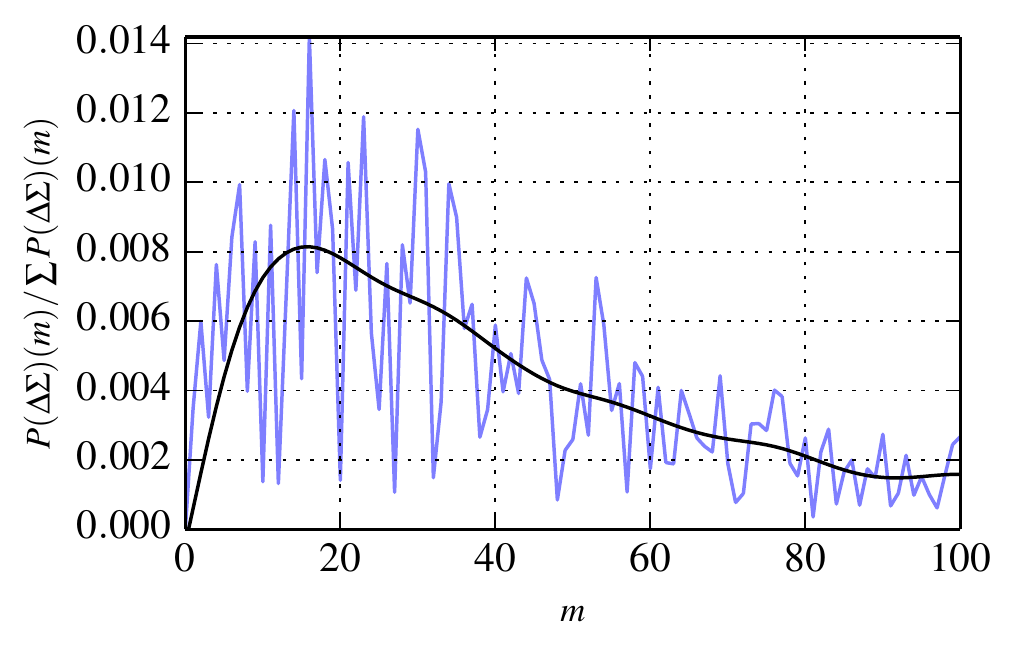}
\includegraphics[width=70mm]{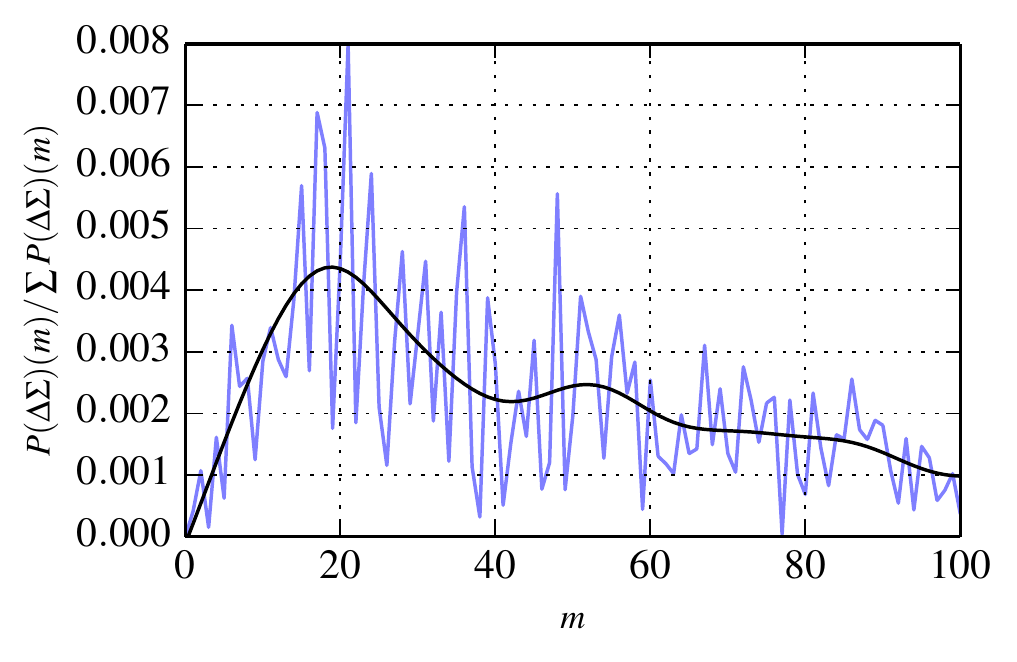}
\includegraphics[width=70mm]{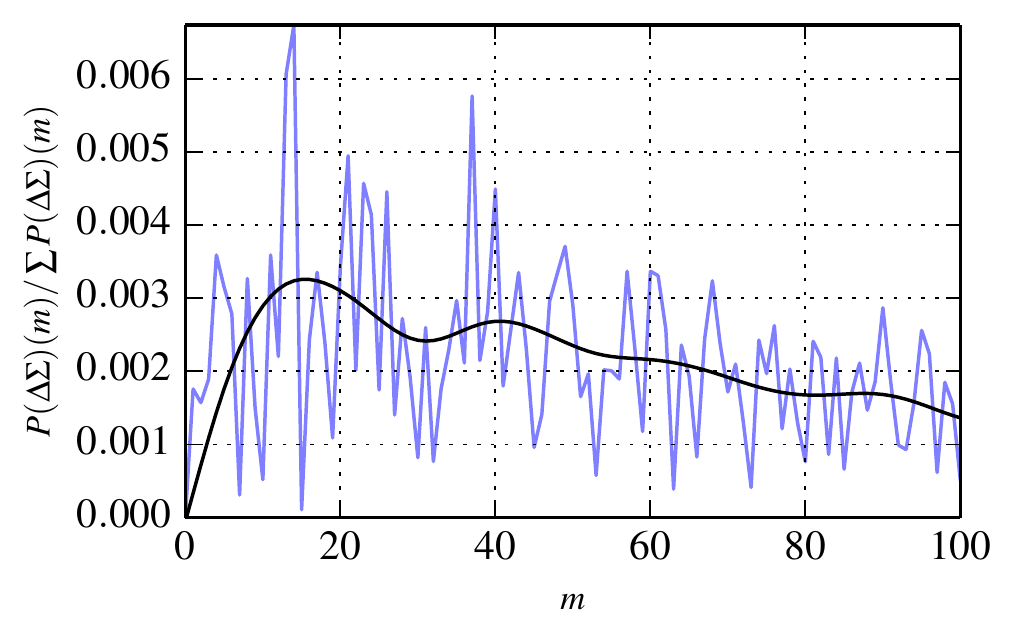}
\caption{Angular spectra of the surface-density with increasing
  resolution.  The data for models A0, A1, A2 and A3 (left to right,
  top to bottom) are shown for $t\simeq 1060~\rm years$.  }
\label{fig:spectra}
\end{center}
\end{figure*}

\begin{figure*}
\begin{center}
\includegraphics[width=0.9\textwidth]{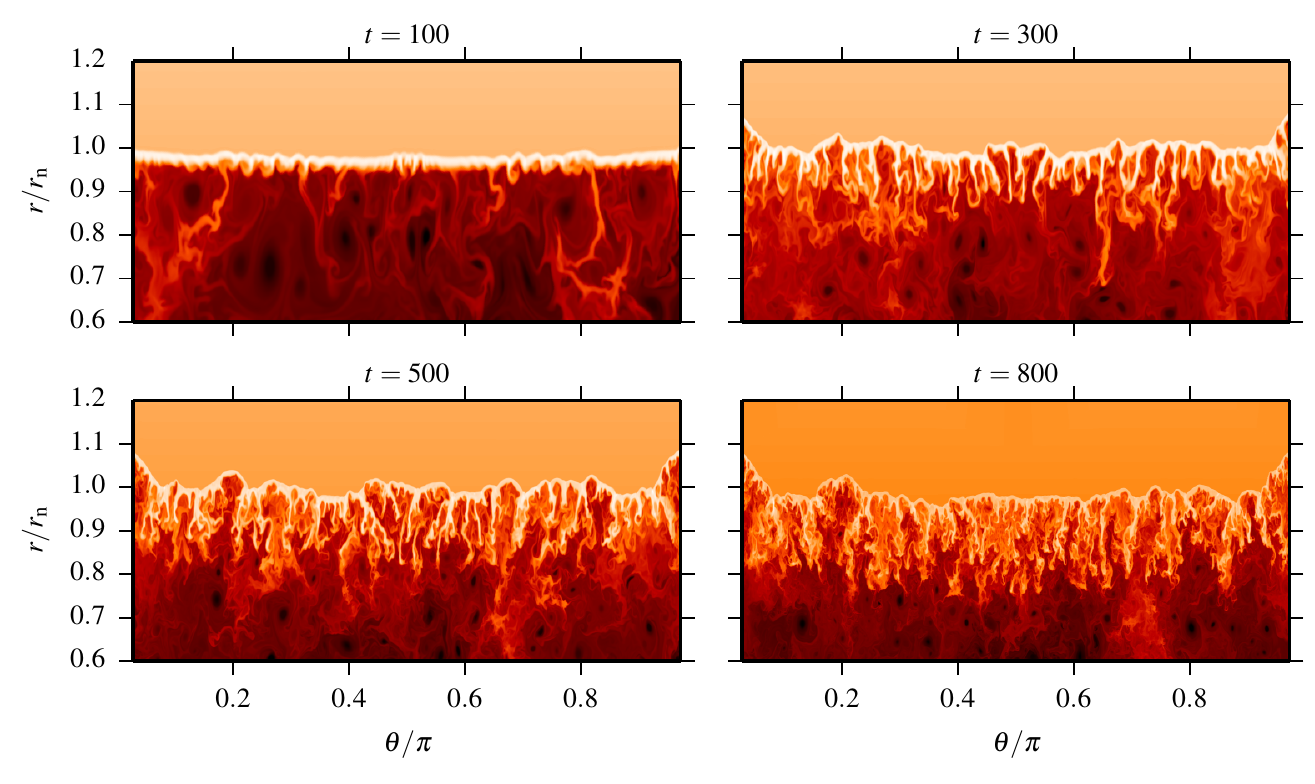}
\caption{Temporal evolution of the filaments for simulation A3 in the self-similar frame.  
We show the quantity $\log_{10}r^2\rho$ for the simulation times $t\in\{100,300,500,800\}$.
The non-linear evolution starting at $t\approx100$ is governed by the merging of established 
filaments to form larger scales as well as fragmentation of bubbles and filaments, 
constantly re-filling fast-growing small scale structure. 
The thickness of the mixing layer saturates at $\sim 20\%$ of the nebula radius.}
\label{fig:time_evol}
\end{center}
\end{figure*}

The results 
are shown in figure~\ref{fig:spectra} together with the low-pass-filtered data.  
They confirm our naked eye observation of the power transfer to 
smaller scale features with increasing resolution. In addition, one can 
see that in all models the spectrum peaks around  $m = 10-20$.
A secondary peak seems to appear at $m\sim 50$ in the model A2 and move 
to $m\sim 40$ in the model A3. 

The growing power of small scales with numerical resolution can be interpreted as 
a result of weaker dampening of small scale RT perturbations by numerical viscosity. 
On the other hand, visual inspection of plots in figure~\ref{fig:resolution} 
also shows that at higher resolution the size of eddies reaching  
the RT interface is also reduced, via development of the turbulent cascade. 
This could be an additional factor in favor of small scale RT modes, as the initial 
perturbations imparted on the RT shell at large scales become weaker, and so require
more time to reach the non-linear regime.  
Moreover, smaller scale eddies are also less powerful and smaller scale 
RT-fingers can survive interactions with them.

Figure~\ref{fig:time_evol} illustrates the time evolution of the RT mixing layer 
in the highest resolution model A3. In order to interpret the data correctly, one 
has to recall that due to the fixed linear resolution, the angular resolution increases
in time following the increase of the linear size of the nebula. This complicates 
the matter. The time $t=100$ plot shows relatively small scale perturbations in the thin dense layer of shocked ejecta (at $r/r_n\sim 1$) reaching the saturation regime. 
This plot also shows much longer and less dense structures curling around the PWN eddies 
in the region $0.6\!<\!r/r_n\!<\!1$. These features are likely to be the result of entrainment 
of the shell matter by the fast flow inside the PWN bubble and not RT fingers. Such 
features have been observed in earlier low-resolution 2D simulations, e.g. the very long 
``fingers'' associated with the backflow of polar jets 
(see figure~6 in \citet{ssk-lyub-04}). At $t=300$, the RT-fingers proper are becoming 
more prominent. They are much longer and occupy the region $0.8\!<r\!/r_n\!<\!1$. 
The angular scale of the shock ripples is also noticeably higher. 

\begin{figure}
\begin{center}
\includegraphics[width=0.45\textwidth]{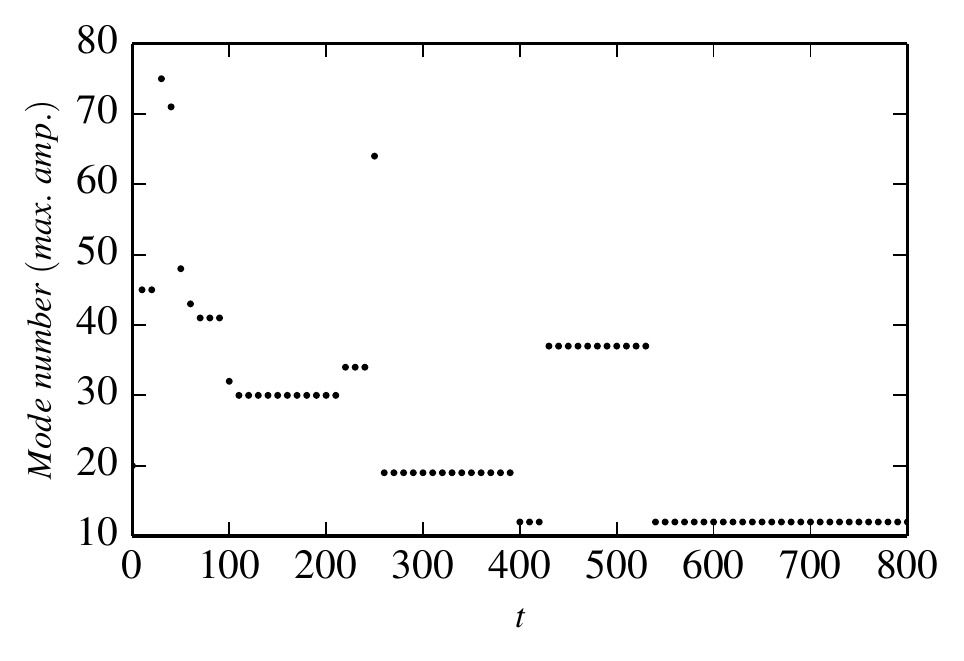}
\caption{Mode number of the most massive scale as a function of simulation time in run A3.  
}
\label{fig:maxamp_vhr}
\end{center}
\end{figure}

At $t=500$ and 800, one can see the fragmentation of large scale shock ripples and large filaments, 
facilitated by the higher effective Reynolds number of the expanding system.  
The increase in Reynolds number can also be seen in the progressively smaller eddies in the PWN proper.  
Fragmentation and inverse cascade of the non-linear RTI compete over the dominant scale of filaments and 
it is not obvious which process has the upper hand at any given time.  
This is visualised in figure~\ref{fig:maxamp_vhr}, showing the time-evolution of the scale containing the most mass.  
\FloatBarrier
Initially, the dominant mode number is small-scale, owing to the faster linear growth of small scale structure.  
The inverse cascade is obtained in the non-linear phase where the larger scales overtake the saturated small scales.  
However, this trend is reversed at times $t\simeq200$ and $t\simeq400$ where we observe a sudden increase 
in the dominating mode number, owing to the creation of new small scale features.  

\begin{figure}
\begin{center}
\includegraphics[width=0.45\textwidth]{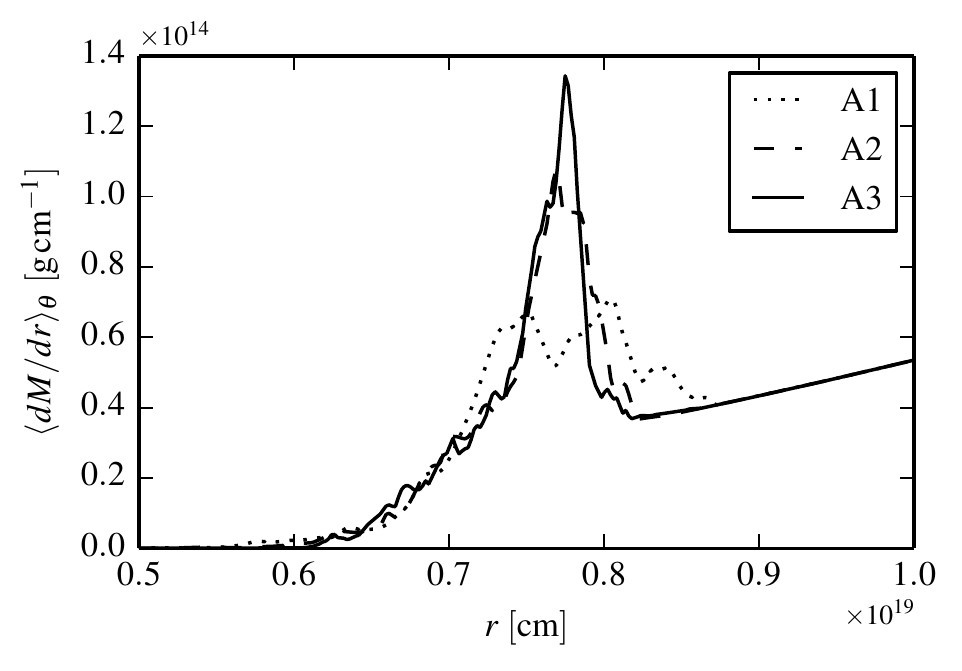}
\caption{Radial distribution of SNR material at $t\simeq950\ \rm years$ for simulation runs \{A1,A2,A3\}.  
}
\label{fig:dmdr}
\end{center}
\end{figure}

While the structure of the filaments in $\theta$-direction thus shows some resolution-dependence, the resulting transport of SNR material into the PWN is largely unaffected by resolution effects.  Figure \ref{fig:dmdr} shows the radial distribution of SNR material defined via
\begin{align}
  \langle dM/dr\rangle_\theta = \frac{1}{\pi} \int_0^\pi 2\pi r^2 \rho_{\rm SNR} \sin\theta d\theta \,.
\end{align}
At $t\simeq 950\,\rm years$, the mixing region ranges from $6\times10^{18} \rm cm$ to $\sim7.5\times10^{18}\rm cm$ and the radial distribution agrees particularly well in the inner part.  Further outside, the distribution becomes increasingly peaked for higher resolution, in agreement with the increasingly circular appearance of the nebula.  In front of the PWN-shock, the SNR evolves according to the self-similar expansion law $\rho_{\rm SNR}\propto t^{-3}$ in good agreement with the simulations.  

\begin{figure*}
\begin{center}
\includegraphics[height=70mm]{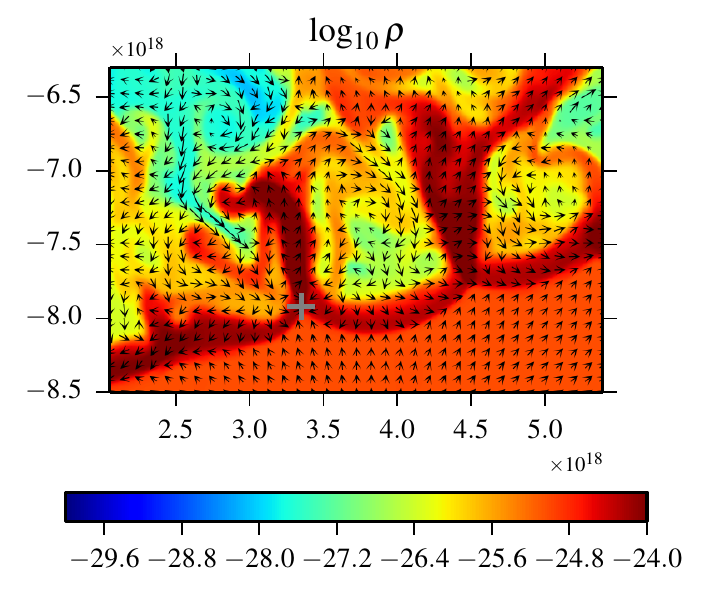}
\includegraphics[height=70mm]{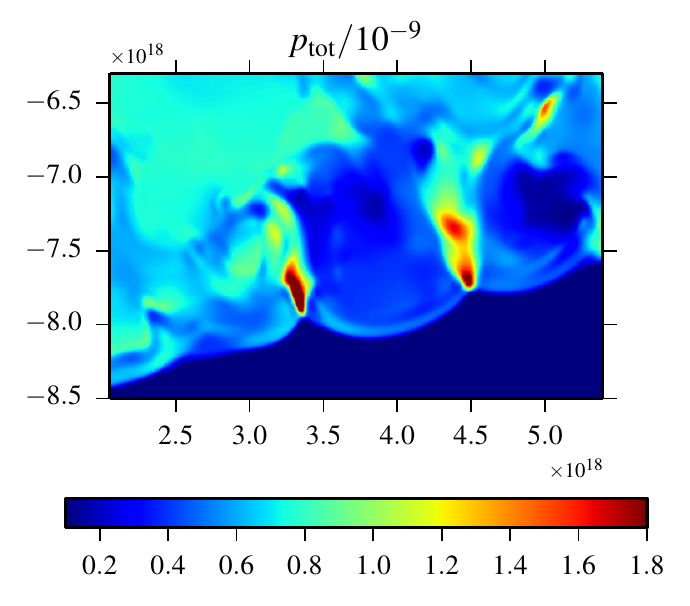}
\includegraphics[height=70mm]{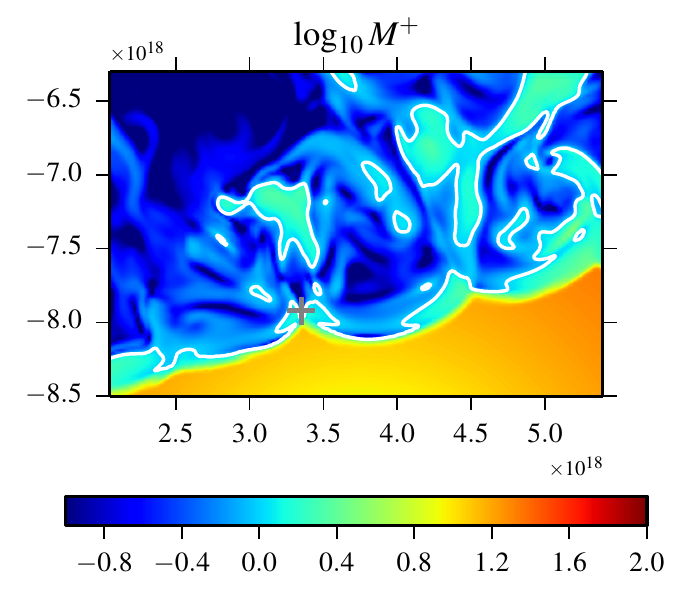}
\includegraphics[height=70mm]{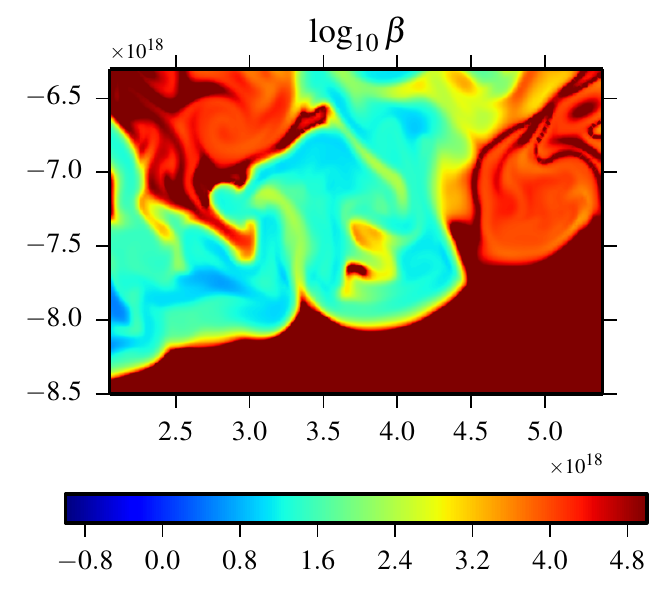}
\caption{Zoom into the southern filaments of run A1
  ($t\simeq1060\rm~years$).  The top-left panel shows the logarithmic
  rest-frame density with flow field vectors and the top-right panel the total pressure.
  Sonic Mach-number with white contour on $M^+=1$ is shown in lower left and
  the ratio of gas and magnetic pressures, $\beta=P_g/P_m$, is given in 
  the lower right panel.
  Note that velocity vectors and Mach number $M^+$ are shown relative
  to the rest-frame of the pivot-point indicated by grey ``+''.  }
\label{fig:zooms}
\end{center}
\end{figure*}

Visual inspection of the animated data (see the online material) suggests that the inverse cascade 
is also present in the simulations -- occasionally smaller scale ripples merge 
and create a larger one. When this occurs, one can see several RT fingers 
emerging from the same base.  The plot for the A3 model in figure~\ref{fig:zoom-rho}
shows an example of such a structure (its base coordinates are $r=3.4$, $z=-8.0$). 
Figure~\ref{fig:zooms} provides more information on the stucture of simpler 
configurations, where only one or two fingers are found at the junction of two 
shock ripples. Interestingly, the total pressure measured at the base of some fingers is 
significantly higher than in the surrounding plasma, with a sharp rise, 
characteristic of a shock wave. In order to check the shock interpretation,     
we studied the velocity field in the frame moving with the velocity measured 
at the base of the left finger, indicated in the figure by a cross.  
This velocity is subtracted from the velocity field measured in the 
original lab-frame and the result is presented in the figure~\ref{fig:zooms}.
\footnote{Since the velocities at the PWN boundary are $\sim2000\rm km\, s^{-1}\ll c$, this Galilei transformation is sufficient.}
This allows us to see clearly the flow converging towards the finger base.  
The Mach number is above unity upstream of the base and drops below 
unity downstream, inside of the high pressure region. Thus, the results 
are consistent with the shocked ejecta plasma sliding with slightly 
supersonic speed along the ripples towards their junction point, 
where it passes through two stationary shocks before entering the 
finger. Vortical motion in the space between the neighbouring fingers
may also contribute to the finger overpressure.

The plasma-beta at the filament base varies greatly (lower right panel), with minimal values larger than $\simeq10$ and maximal values -- in regions where mixing is advanced -- ranging up to $10^4$.  
In simulation B1 with $100\times$ stronger wind magnetisation, with the exception of the spurious jet region where $0.1<\beta<1$, we still observe similar values of the interface plasma beta, due to annihilation of flux loops with opposite polarity in the nebula.  
Thus we do not expect strong suppression of field-aligned modes, as the critical angular scale becomes $\theta_c/\pi\lesssim 1/40$  ( see Eq. \ref{eq:theta_c}) in the bulk of the nebula.

\section{Discussion}
\label{sec:discussion}

The convergence study of the RT-instability described in the previous 
section indicates that with higher resolution, the filamentary structure 
produced in the simulations becomes less similar to the one observed in 
the Crab Nebula, with the deficit of large-scale structure being 
the most pronounced discrepancy. In this section, we discuss the possible 
explanations of this result and speculate on the origin of the Crab's 
large-scale filaments.        

As we have already commented on in the Introduction, the radial expansion of 
the RT interface in our problem, introduces a number of interesting modifications
to the classical results obtained in plane geometry.   
Here we derive them via adjusting in a rather crude but very simple way 
the non-relativistic results in planar symmetry. More accurate analysis of the linear 
theory can be found elsewhere \citep{BT-53,GL-86,CF-92,goedbloed2010}.  
We start with the non-magnetic case and discuss the potential role of 
magnetic field later.

We denote as $\theta_m=\lambda/r_n$ the angular scale of RT perturbations. 
Here $r_n\propto t^\delta$, where $\delta=6/5$, is the
mean radius of the nebula at the phase of self-similar expansion. 
As the nebula expands radially, $\lambda$ increases proportionally to $r_n$, 
but $\theta_m$ remains constant. This linear stretching of the perturbations 
is most important. First, the linear amplitude of perturbations, 
$h$, is no longer the most suitable parameter to describe their strength  
and should be replaced with $x=h/\lambda$. 
Assuming that $h$ grows at the same rate as in the plane case with $A=1$ we find 

\begin{align}
 \dot{x}=x\left[ \frac{\dot{h}}{h} - \frac{\dot{r}_n}{r_n}\right] = 
  \kappa \frac{x}{t}\, ,
\label{eq:x}
\end{align}
where
$$
   \kappa=\omega - \delta = \fracp{2\pi\delta(\delta-1)}{\theta_m}^{1/2}-\delta \,.
$$
This shows that the amplitude growth is a power law, $x\propto t^\kappa$,
and suggests the critical angular scale 
$$ 
\theta_s = \frac{2\pi(\delta-1)}{\delta} 
$$ 
above which the perturbations do not grow. For $\delta=6/5$, corresponding to 
a uniform ejecta and constant wind power, $\theta_s=\pi/3$. More accurate 
analysis of RTI in spherical geometry yields the growth rate 
\begin{align}
  \omega^2 = \frac{lg}{r_n} \, ,
\label{eq:omega_sph}
\end{align}
where $l$ is the degree of the associated Legendre polynomial $P^m_l(\cos\theta)$ 
\citep{BT-53,GL-86}. In the limit of high $l$ we should recover the plane 
geometry which leads to the identification of the wavenumer $k$ with $l$ via $k=l/r_n$. 
Using this we find the critical degree $l_s=6$ which corresponds to $\theta_s=\pi/3$. 
This is very close to the critical degree $l_s=5$ found in the thin shell 
approximation \citep{CF-92}, showing that the dumping of the RT instability at 
large scales is a robust result.  Based on this we tentatively 
conclude that there is an upper limit of $\simeq 50^\circ$ for the angular size of 
perturbations above which they do not grow.     

The solution to Eq.~(\ref{eq:x}), 
$ x(t)=x_0 (t/t_0)^\kappa $, implies that a perturbation of any amplitude imposed 
at the time $t_0$ will be able to reach the non-linear regime provided $t_0$ 
is sufficiently small. For the Crab Nebula this means that we should 
be dealing with the fully nonlinear regime on all scales. However, in our simulations where $t/t_0$ is not particularly high, some large-scale perturbations 
may still be growing in the linear regime. 
For example, according to this result, the largest angular scale to grow by a 
factor of $e$ during our simulation time is $\pi/4.4$. 
Moreover, one may reasonably expect 
their final amplitude to depend on the strength of large-scale motion inside the PWN
bubble as it is responsible for the initial amplitude of these perturbations. 
The reduction of the final amplitude with resolution, observed in our simulations, 
may well reflect the parallel weakening of this large scale motion.    

Since in the real Crab Nebula the perturbations of all scales which are linearly unstable, are 
expected to have reached the nonlinear regime, this regime is much more relevant for 
interpretating the observations. To this end,   
consider the growth of RT bubbles in the non-linear regime. Substituting 
$g=\ddot{r}_n=\delta(\delta-1)(r_n/t^2)$ into equation~(\ref{eq:v_b}) and integrating, 
we find that the bubble height is 

\begin{align}
   \frac{h}{r_n} \simeq \frac{1}{2}\fracp{\theta_m (\delta-1)}{\delta}^{1/2} \, .
\label{eq:h_b}
\end{align}
Thus, the relative height of bubbles, $h/r_n$, ``freezes out'' in the non-linear
phase.  This conclusion fits nicely the picture of self-similar expansion. 
The critical scale $\theta_s$ of linear regime sets the upper limit on $h/r_n$. 
For $\delta=6/5$ this limit is $(h/r_n)_{\rm max}\simeq 0.2$, which is about the size 
for the largest bubbles of the Crab Nebula ``skin'' \citep{hester-08}.  
The fact that in our simulations we do not observe  high amplitude bubbles 
on these scales may indicate they have not had enough time to reach the non-linear 
regime yet. The inverse cascade may contribute to the production of large-scale 
bubbles, but it is unlikely to overturn the freezing-out effect. 

Since the finite 
thickness of the RT unstable layer and the forward shock do not feature in the 
analyses leading to Eq.~(\ref{eq:h_b}), this result should not be considered as an 
accurate prediction yet. However, it gives us a basis to speculate about the 
origin of the largest ``filaments'' seen in the Crab Nebula. These features 
can be as long as the nebula radius and they do not appear to be streaming 
radially towards its center (see figure 1 in \citet{hester-08} as well as 
figure 2 in \citet{clark-83})  as one would expect for the RT-fingers. Instead 
they seem to outline a network of very large cells filled with synchrotron-emitting 
plasma.  We propose that these cells are actually the largest RT-ripples (bubbles) 
on the surface of the Crab Nebula and these filaments designate  ``valleys'', 
where these ripples come into contact with each other. The plasma of shocked ejecta 
may slide along the surface of the ripples into these valleys, in very much the same 
fashion as we have discussed in connection with figure~\ref{fig:zooms}.  
These filaments may form a base from which proper RT-fingers will stream 
radially towards the center of the nebula. In fact, this is indeed 
what is seen in the Crab Nebula, most 
clearly in its NE section, where a number of smaller scale filaments seem to 
originate from a large one at the angle of almost 90 degrees. Remarkably, the 
observed cell size is in a very good agreement with the largest angular 
scale of shock ripples, $\theta_s\simeq \pi/3$, which can be amplified by the 
RT-instability. These large-scale ripples
are not seeded internally by the interaction with the large-scale motion inside the PWN 
bubble, the only source of perturbations in our simulations.  
Instead, they may originate from inhomogeneities in the supernova ejecta itself, 
which we did not incorporate in our models. Given the violent nature of supernova 
explosions it seems only natural to expect strong large-scale fluctuations in the 
ejecta \citep[e.g.][]{CO-13}. Moreover, 
\citet{FMS-92} argued  that the conspicuous ``bays'' in the nonthermal 
optical emission of the Crab Nebula could be indications of a presupernova 
disk-like ejection. The interaction of a supernova ejecta with such a disk is 
believed to be behind the emergence of bright rings around 
SN 1987 A \citep{larsson-11}.  

As we have noted in the Introduction, the magnetic field may have strong 
impact on the development of the RT-instability. This may seem particularly 
significant as pulsar winds inject highly magnetized plasma into the PWN 
bubble. Even for weakly-magnetised winds, the magnetic effects would be 
important provided PWN were organised in accordance with the Kennel-Coroniti 
model. In this model, the initially weak magnetic field is amplified towards 
equipartition between the magnetic and thermal energies near the 
contact discontinuity with the shocked supernova ejecta. This is exactly 
the condition for inhibiting the RTI as derived in \citet{bucc-04}, when 
considering modes aligned with the magnetic field.  
In contrast, in our simulations the magnetic field is always normal to the 
wave vector of any type of perturbation due to their symmetry, which nullifies
the magnetic effect and which can be considered as the main limitation of our 
study. However, this limitation is probably not as important as it seems. 
Strong magnetic dissipation seen in our simulations, particularly 
in the 3D models \citep{porth-13,porth-14}, keeps the magnetic field well below 
the equipartition near the interface even for high-sigma pulsar winds.  
Moreover, in 3D the magnetic field is not that effective in inhibiting the RTI, 
as matter can slide in between the magnetic field lines without bending them 
downwards \citep{SG-07}. The combination of these factors makes us
conclude that the impact of magnetic field on the development of RTI in 
the Crab Nebula is likely to be rather minimal, which is consistent 
with the observations.

Apart from the high magnetization employed in some runs, 
the setup of our simulations is very similar to that of the previous 
axisymmetric simulations of PWN, e.g. by \cite{ssk-lyub-03,ssk-lyub-04,
delzanna-04,bogovalov-05,camus-09}.  However, none of those captured
the development of the RT-instability.  We believe that this is due to
the insufficient resolution of previous studies at the interface
between the PWN and the supernova shell.  This is not surprising as
these studies were mainly concerned with the inner regions around the
termination shock and used spherical coordinates which fit the purpose nicely. 
However, their spatial resolution thus quickly decreases with the distance 
from the origin.  In contrast, in the
cylindrical coordinates employed here, we obtain uniform resolution
throughout the PWN ($\Delta r=\Delta z =1.95\times 10^{16}\, \rm cm$).
Moreover, we utilize third order spatial reconstruction and Runge-Kutta 
time-stepping giving overall higher accuracy compared to the previous 
studies.

\section{Conclusions}
\label{sec:conclusions}

Our high resolution axisymmetric simulations of PWN now reveal intricate structures of filaments growing via the Rayleigh-Taylor instability of the contact discontinuity between PWN and SNR.  
Given the high rate of magnetic dissipation observed in recent 3D simulations of PWN, the magnetic tension is likely to play only a minor role in the RTI such that our axisymmetric simulations are in fact applicable to reality.  

In application to the Crab nebula, we have simulated the last 800 years of its evolution and find the longest fingers to reach a length of $\sim1/4$ of the nebula radius.  
The inverse cascade observed in the planar RTI is complemented by constant replenishment of small scale structure due to fragmentation of old filaments and formation of new fast growing small scale perturbations.  
The latter is particularly pronounced at the large ``bubbles'' of the nebula which, as they expand along with the nebula provide favourable conditions for growth of fresh small-scale RTI.  

The most massive filaments in our simulations reach a scale of $m\simeq15$ (corresponding to 15 large fingers over the semi-circle), independent of the numerical resolution.  
Our simulations can not yet reproduce the largest scales observed in the Crab nebula and we propose that they must be seeded from inhomogeneities in the SNR as would result from an anisotropic supernova explosion.  

In the future, we plan to investigate the influence of magnetic tension on the filamentary network in local 3D simulations of PWN with realistic values of the magnetisation.  

\section{Acknowledgments}
SSK and OP are supported by STFC under the standard grant
ST/I001816/1.  SSK has been partially supported by NASA grant NNX13AC59G.  
SSK acknowledges support by the Russian Ministry of
Education and Research under the state contract 14.B37.21.0915 for
Federal Target-Oriented Program.  RK acknowledges FWO-Vlaanderen,
grant G.0238.12, and BOF F+ financing related to EC FP7/2007-2013
grant agreement SWIFF (no.263340) and the Interuniversity Attraction
Poles Programme initiated by the Belgian Space Science Policy Office
(IAP P7/08 CHARM). The simulations were carried out on the Arc-1
cluster of the University of Leeds.

\bibliographystyle{mn2e}
\bibliography{pwn,mypapers,lyubarsky,lyutikov,hea,plasma,numerics,mix,astro}

\end{document}